\documentclass[pre,twocolumn,superscriptaddress,aps,notitlepage,10pt]{revtex4-1}

\usepackage{natbib}

\usepackage{times}
\usepackage{amssymb}
\usepackage{amsmath}
\usepackage{bm}
\usepackage{color}
\usepackage[colorlinks=true,urlcolor=blue,breaklinks,citecolor=blue]{hyperref}
\usepackage{graphicx}
\usepackage{braket}
\usepackage{subfigure}
\usepackage{xcolor}

\graphicspath{{.}{Images_Paper/}}

\begin{document}

\title{Automated discovery of characteristic features of phase transitions in many-body localization}

\author{Patrick Huembeli}
\affiliation{ICFO-Institut  de  Ciencies  Fotoniques,  The  Barcelona  Institute  of Science  and  Technology, 08860  Castelldefels  (Barcelona),  Spain}
\author{Alexandre Dauphin}
\affiliation{ICFO-Institut  de  Ciencies  Fotoniques,  The  Barcelona  Institute  of Science  and  Technology, 08860  Castelldefels  (Barcelona),  Spain}
\author{Peter Wittek}
\affiliation{University of Toronto, M5S 3E6 Toronto, Canada}
\affiliation{Creative Destruction Lab, M5S 3E6 Toronto, Canada}
\affiliation{Vector Institute for Artificial Intelligence, M5G 1M1 Toronto, Canada}
\author{Christian Gogolin}
\affiliation{ICFO-Institut  de  Ciencies  Fotoniques,  The  Barcelona  Institute  of Science  and  Technology, 08860  Castelldefels  (Barcelona),  Spain}
\affiliation{Institute for Theoretical Physics, University of Cologne, 50937 K\"oln, Germany}
\affiliation{Xanadu, 372 Richmond St W, Toronto, M5V 1X6, Canada}

\begin{abstract}
  We identify a new ``order parameter'' for the disorder driven many-body localization (MBL) transition by leveraging \textcolor{black}{machine learning}.
  Contrary to previous studies, our method is almost entirely unsupervised.
  A game theoretic process between neural networks defines an adversarial setup with conflicting objectives to identify what characteristic features to base efficient predictions on.
  This reduces the numerical effort for mapping out the phase diagram by a factor of ~100x and allows us to pin down the transition, as the point at which the physics changes qualitatively, in an objective and cleaner way than is possible with the existing zoo of quantities.
  Our approach of automated discovery is applicable specifically to poorly understood phase transitions and is a starting point for a research program leveraging the potential of machine learning assisted research in physics.
\end{abstract}

\maketitle

\section{Introduction}

Can \textcolor{black}{machine learning} (ML) offer a qualitative advantage by assisting scientific discovery?
Or is it just a new tool for numerical calculations?
Machine learning has been making headlines in computational physics: it proved remarkably efficient in giving comparable accuracy to known methods for the study of phase transitions~\cite{nieuwenburg2017learning,carrasquilla2017machine,Chng2017,wetzel2017unsupervised,wang2016discovering,zhang2018machine, schindler2017probing,Hu2017, broecker2017}.
Here, we show that state-of-the-art ML is capable of more, by automating the discovery of robust characteristic features, which will enable a more efficient investigation of physical effects.

An example where \textcolor{black}{ML} assistance is much needed is the delineation and characterization of the many-body localized (MBL) phase, exhibited by systems with many interacting quantum particles experiencing a (strong enough) static disordered background potential.
This research problem has attracted an immense amount of attention recently \cite{Basko2006d,Pal2010,Nandkishore2014,Luitz2015,Thiery2017,Parameswaran2018,Pietracaprina2018,Abanin2018} because MBL challenges long-held believes about the phase structure of isolated systems and even the applicability of standard equilibrium statistical mechanics, which no longer correctly captures the long-time behavior in that phase.
Many details of how this breakdown happens remain elusive, despite the known characterization of MBL in terms of local conserved quantities \cite{Serbyn2013,Huse2014} and an extensive and ongoing debate \cite{Nandkishore2014,Parameswaran2018,Thiery2017,Abanin2018}.
Our work addresses two major roadblocks preventing further progress:
First, it yet remains unclear what the best approach is to delineate the MBL phase.
Physicists have come up with a whole zoo of quantities whose behavior can serve as an indicator for the transition, but the various phase boundaries they imply, do not agree within error bars \cite{Luitz2015} and controlling finite-size effects is a challenge \cite{Kjaell2014,Enss2017}.
Second, all of these quantities need to be averaged over an enormous number of disorder realizations (often 10.000 \cite{Pal2010,Kjaell2014,Luitz2015,Kudo2018}) to get meaningful results.
Highly optimized codes allow in principle to study systems of up to 26 spins \cite{Pietracaprina2018}, but with the known quantities, going beyond 22 spins is prohibitively expensive because of disorder averaging \cite{Pietracaprina2018}.

\begin{figure}[tb]
  \includegraphics[width=\linewidth]{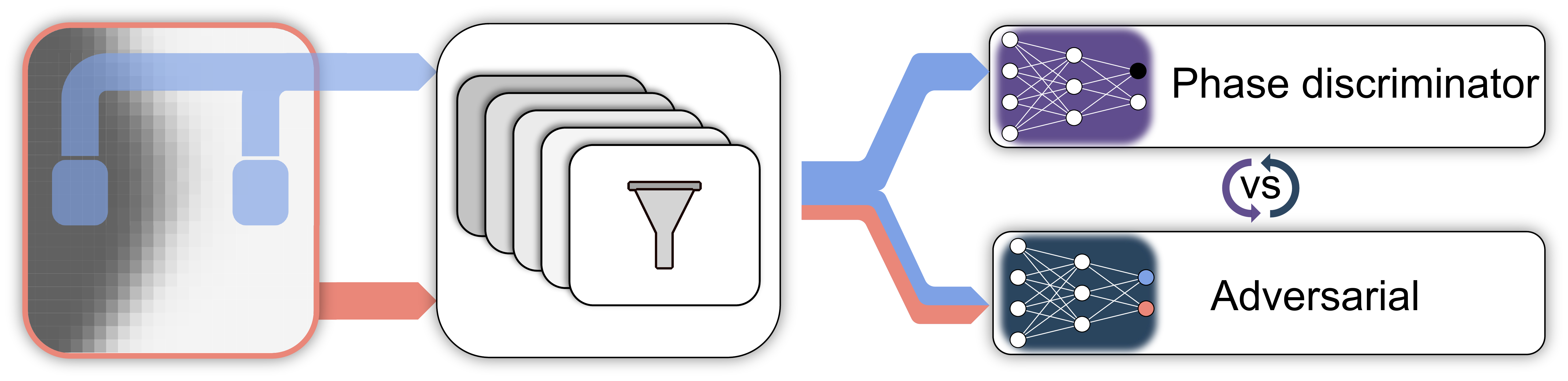}
  \caption{\label{fig:DANN_overview}
    By using a contemporary neural network architecture, we automate feature extraction and drastically reduce computational cost at the same time.
    To achieve this higher level of automation, a pair of neural networks share a pipeline for feature extraction. They compete in a game theoretic framework to achieve conflicting goals:
    one network has to classify states according to their phase and the other is supposed to tell from how deep in the phase they are.
    The equilibrium of the game tells what features to base predictions on and allows to determine the phase boundary in a largely unsupervised way.
  }
\end{figure}

Through ML assisted research, we obtain a quantity capturing features of the MBL transition that is so powerful that up to 100x fewer disorder realizations are sufficient to obtain objective and more accurate predictions of the transition point.
We thereby reduce the cognitive load of the scientist by automating feature extraction by means of a game theoretical process in the state-of-the-art adversarial domain adaptation technique (Figure~\ref{fig:DANN_overview}).

\section{Automated feature extraction.}
If the phases are well understood, standard supervised deep learning can be used to find out which phase a test state comes from \cite{carrasquilla2017machine,Chng2017,zhang2018machine}.
 \textcolor{black}{ Unsupervised techniques have so-far been used in classical systems \cite{wetzel2017unsupervised,wang2016discovering} or rely on the knowledge that manually engineered features, such as the entanglement spectrum or spin-spin correlators, capture the physics of the phase transition \cite{nieuwenburg2017learning, broecker2017}.}

The problem we want to solve here, however, is qualitatively different.
We want to automatically learn the unknown location of a phase boundary in a largely unsupervised way and without engineering features by hand in a phase transition of a quantum model that remains a challenge for existing methods.

We achieve this by means of adversarial domain adaptation \cite{ganin2016domain,huemb2018}, a technique in which two neural networks are competing in a game.
The networks share a common feature extraction pipeline that consists of convolutional and pooling layers as in ordinary deep learning (Figure~\ref{fig:DANN_overview}).
Two types of input data are fed into the shared pipeline.
The first type has labels.
For instance, we can easily select states deep inside the phases and confidently label them.
The second type of data contains points for which the label is unknown.
This can be states from all over the phase space, including, in particular, such from around the suspected position of the phase boundary.

The first neural network receives only the first type of data.
Its goal is to maximize prediction accuracy of the label.
If the architecture did not have more components, this would be a similar scenario to the ones discussed in previous work~\cite{nieuwenburg2017learning,carrasquilla2017machine,Chng2017,wetzel2017unsupervised,zhang2018machine, schindler2017probing}.
The key difference is that this first network has an adversary, who receives both types of data and is tasked with guessing whether a data point is labeled or not.
The common feature extraction pipeline is adjusted to make the task of the first network as easy as possible while making that of the second as hard as possible.
This is achieved by means of error backpropagation from both networks, but with opposite signs.
When the game reaches equilibrium, the representation layer selects features that are best suitable to identify which phase a state comes from, but contain virtually no information about from where inside the phase they are.

More formally, the phase discriminator is endowed with a loss function $L_d$ that it tries to minimize.
This function depends on two sets of parameters: $\theta_d$, which are the parameters describing only the phase discriminator neural network, and $\theta_f$, which are the parameters describing the feature extraction layers.
The loss function $L_a$ of the adversary is a function of $\theta_f$ and the parameters $\theta_a$ describing this network alone.
The game is about achieving an equilibrium in $\theta_f$, through the update rule $\Delta\theta_f = \mu (\frac{\partial L_d}{\partial \theta_f}-\frac{\partial L_a}{\partial \theta_f})$, where $\mu$ is the learning rate.
The opposite sign in the gradient update expresses the competition between the networks (for more details see in the Appendix \ref{sec:domain_adversarial_neural_network}).

This competitive process enables the learning algorithm to autonomously figure out a (possibly non-local) ``order parameter'', pinpointing where the physics changes qualitatively.
The last layer of the phase predictor gives the probability that an input state is part of one or the other phase.
We can then discard the adversary and use only the output of the first neural network to predict all labels.
To decrease the noise close to the phase boundary this output can be averaged over several disorder realizations.
The learning is completed and when the predicted labels no longer change.

\section{Results}
We apply the adversarial network architecture to the problem of delineating the MBL phase boundary in the prototypical spin-$1/2$ Heisenberg chain in a random magnetic field, described by the Hamiltonian

\begin{equation} \label{eq:heisenberg_chain_hamiltonian}
  H = \frac{1}{2} \, \sum_{i = 1}^N \sum_{\alpha \in \{x,y,z\}} \sigma^{\alpha}_i \, \sigma^{\alpha}_{i+1} - \sum_{i = 1}^N h_i \, \sigma_i^z, 
\end{equation}
with $\sigma^{x,y,z}_i$ the Pauli matrices on site $i$ and the $h_i$ are drawn from the uniform distribution over $[-h,h]$.
We denote the normalized energy by $\epsilon \in [0,1]$, which interpolates between the lowest and highest of the energies of $H$ for a given realization of the disordered fields $h_i$ and restrict to the global magnetization zero subspace.
The eigenstates of this model are known to undergo an MBL transition at an energy dependent critical disorder strength $h_c$, whose precise position is however difficult to determine with established methods.
The most widely used method to detect the MBL transition is the average adjacent gap ratio $r$ \cite{Luitz2015,Kudo2018}, which goes from $r_{\text{WD}} \approx 0.53$, resulting from the Wigner-Dyson distributed eigenvalues in the ergodic phase, to $r_{\text{Poisson}} \approx 0.38$, reflecting the Poisson statistics in the MBL phase.
Another quantity is the dynamical spin fraction $f$, which varies from $1$ to $0$ \cite{Pal2010,Luitz2015}.

We generate eigenstates from small windows around several values of $\epsilon$ and for multiple disorder realizations at different disorder strengths $h$ for system sizes up to $N=18$ spins with the shift invert code from \cite{Pietracaprina2018} (Details in the Appendix~\ref{sec:input_data}).
For the training of the network we use as the first type of data states from two sets deep inside the phases.
For the second type we generate states from a wide range of $\epsilon$ and $h$ values that including the phase boundary (see Figure~\ref{fig:DANN_overview}). We want to emphasize that we use the coefficients of the wavefunction as input data without further preprocessing.

\begin{figure}[tb]
  \subfigure[]{
    \includegraphics[width=0.46\linewidth]{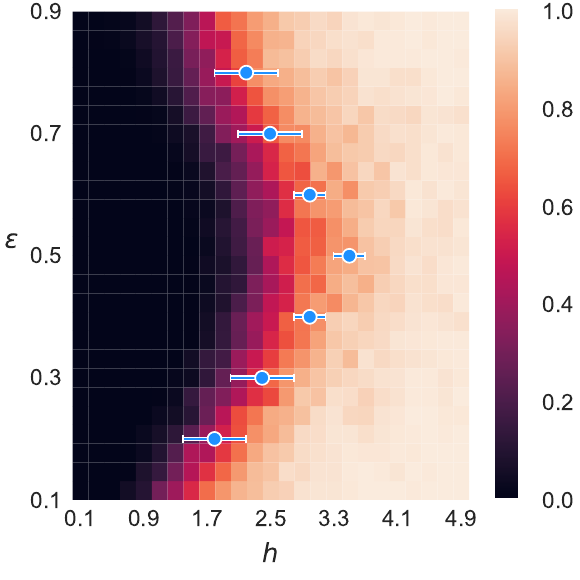}
    \label{fig:2d_plot_nn}
  }
  \subfigure[]{
    \includegraphics[width=0.46\linewidth]{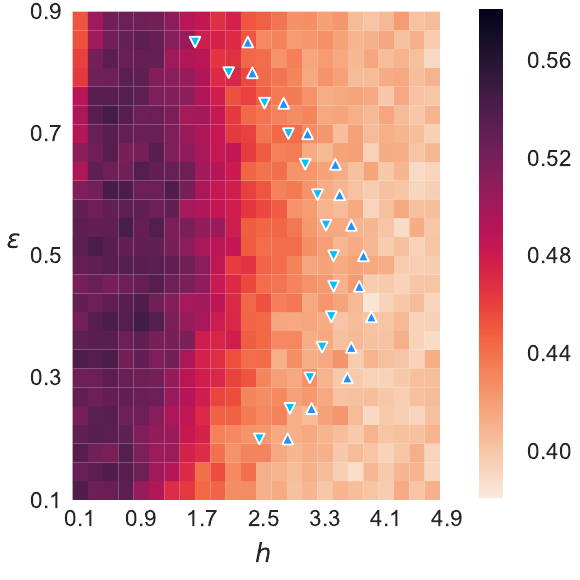}
    \label{fig:2d_plot_gs}
  }
  \caption{\label{fig:2d_coparison_of_nn_to_gap_statistics}
    The output of the neural network directly provides a meaningful estimate of the phase diagram for a finite system $N=12$ (background in \subref{fig:2d_plot_nn}) from just $50$ disorder realizations, while traditional quantities, like the gap statistics (background in \subref{fig:2d_plot_gs}) are still far too noisy.
    The dots in \subref{fig:2d_plot_nn} are the extrapolated phase boundary in the thermodynamic limit obtained from $100$ disorder realizations via the data collapse for systems up to $N=18$, shown exemplary for $\epsilon=0.5$ in Figure~\ref{fig:coparison_of_nn_to_gap_statistics_at_epsilon_one_half}.
    The symbols in \subref{fig:2d_plot_gs} are the phase boundaries found in \cite{Luitz2015} based on the average adjacent gap ratio $r$ (triangles) and the dynamical spin fraction $f$ (triangles pointing downwards) for systems of size up to $N=22$ and vastly more disorder realizations (the data collapse plots for all values of $\epsilon$ are shown in the Appendix in Figure \ref{fig:collapse_for_all_epsilon}).
  }
\end{figure}

We compare the estimate of the energy resolved phase diagram obtained from the adversarial neural network with results based on the average adjacent gap ratio $r$ and the dynamical spin fraction $f$.
The superior statistical properties of our approach are apparent.
Already from only $50$ disorder realizations we obtain a clear characterization of the phases, while the average adjacent gap ratio is still very noisy (background in Figure~\ref{fig:2d_coparison_of_nn_to_gap_statistics}).

\begin{figure}[tb]
  \centering
  \subfigure[]{
    \includegraphics[width=0.46\linewidth]{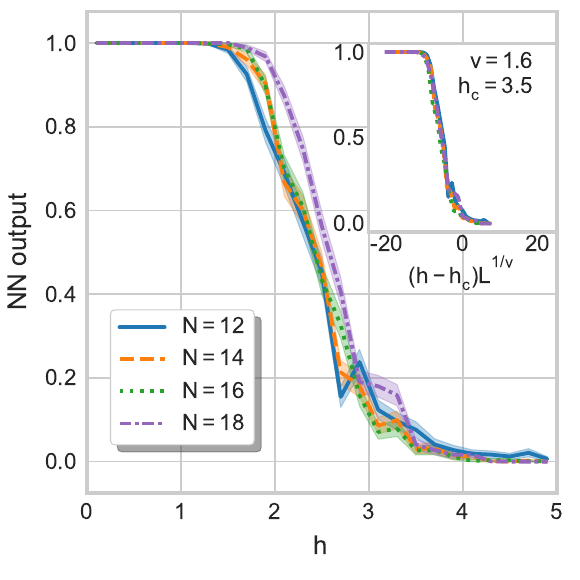}
    \label{fig:nn_output_at_epsilon_one_half}
  }
  \subfigure[]{
    \includegraphics[width=0.46\linewidth]{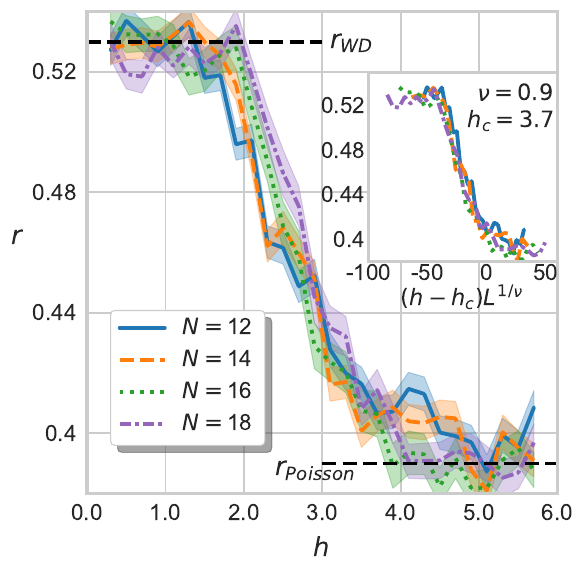}
    \label{fig:gap_statistics_at_epsilon_one_half}
  }
  \caption{\label{fig:coparison_of_nn_to_gap_statistics_at_epsilon_one_half}
    Exemplary output \subref{fig:nn_output_at_epsilon_one_half} of the neural network at normalized energy $\epsilon = 0.5$ averaged over $50$ disorder realizations and the data collapse (inset) to determine the position of the phase transition $h_c$ in the thermodynamic limit.
    The average adjacent gap ratio $r$ is still far too noisy \subref{fig:gap_statistics_at_epsilon_one_half} to obtain a good collapse (inset) for the same amount of averaging.
    The error bands show the ensemble standard deviation $s =  (\sum_i^N (x_i-\hat{x})^2 / (N-1))^{1/2}$ of the disorder average.}
\end{figure}

The phase boundary shown in Figure~\ref{fig:2d_plot_nn} can be determined via a data collapse from plots such as that shown in Figure~\ref{fig:coparison_of_nn_to_gap_statistics_at_epsilon_one_half} for $\epsilon = 0.5$.
To extract the critical magnetic field strength $h_c$, we use a scaling function of the same form $N^{1/\nu}(h-h_c)$ as that for the average adjacent gap ratio $r$.
The reduction in noise allows for a more precise determination of the phase boundary for the same number of disorder realizations.
From just data for systems up to size $N=18$ (100 disorder realizations) we are able to determine the phase boundary extrapolated to the thermodynamic limit to an accuracy roughly matching the discrepancy between the conventional quantities $r$ and $f$ determined in the numerically much more expensive study \cite{Luitz2015}.
Intuitively, it makes sense that the average adjacent gap ratio does not have the nice averaging properties of the quantity computed by our neural network, as it completely disregards the properties of the eigenstates and only computes one feature of the spectrum.
ML, in contrast, figures out a way to objectively determine the phase by directly recognizing non-trivial properties of the eigenstates.  

Another interesting feature in which our method differs from average adjacent gap ratio (as well as most other quantities that have been used in exact diagonalization studies so far) is the value of the scaling exponent $\nu$.
Our method consistently yields $\nu \approx 1.6$, independent of the energy range and the precise choice of the training data (under the condition that it is sufficient to ensure convergence of the training), while the average adjacent gap ratio yields $\nu\approx0.9$ \cite{Luitz2015}.
Both exponents violate the (heuristic) Harris criterion, which for one spacial dimension predicts $\nu>2$ \cite{Harris1974,Harris2016}, but the larger value of our ``order parameter'' is closer to the predicted value and there is hope that by moving to even larger system sizes, the best data collapse will be obtained with $\nu \approx 2$.
This is another indication that our automatically detected ``order parameter'' suffer less from finite size effects than more traditional quantities.
The size of the region in which the network is unsure which label to assigned shrinks during training and eventually converges.
It is a natural measure for the broadening of the phase transition due to finite size effects.

Our method has a number of additional desirable properties.
The intermediate values of the average adjacent gap ratio do not have a physical meaning, whereas the output of the neural network has an immediate interpretation as to how certain the phase prediction is.
The predicted values of $h_c$ and the sizes of the plateaus are stable against changing the regions from which the first kind of training data is generated.
The average adjacent gap ratio, actually attains the Poisson value at the integrable point at $h=0$ and it moreover fails to capture the transition if one does not restrict to a fixed magnetization sector.
Our method does not suffer from either of these two drawbacks.

Importantly, the computational time for training and evaluating the output of the adversarial neural network is almost negligible compared to the time it takes to generate states for mapping out the phase diagram.
As much fewer disorder realizations are necessary per point, this yields a huge net gain in computational time.
Our approach thus will allow to meaningfully include states from larger system sizes, which can now be generated with state-of-the-art shift invert algorithms \cite{Pietracaprina2018}, into studies of MBL.

\section{Conclusions}
We have demonstrated that ML can be used to automate the task of identifying relevant features that most efficiently capture the physics of phase transitions in quantum systems --- a formidable task so far reserved for human researchers.
Concretely, the competitive process of adversarial domain adaptation, is able to ``invent'' a new ``order parameter'' for the MBL phase transition that yields meaningful results from vastly fewer disorder realizations than established methods.

It seems fair to say that the resulting quantity actually captures the essential physics, as the network, once trained, can correctly identify the phase transition not only at different energy densities, but also in similar but distinct models.
This is remarkable, since the MBL transition has mostly defied analytical approaches and even the question of what is the best way to delineate the phase could not be resolved in a satisfactory way.
Our method is directly applicable to other non-standard critical phenomena beyond MBL and can be used to distinguish multiple phases, even across different classes of models, as long as their Hilbert spaces are compatible \cite{huemb2018}.

As the automatic feature identification does not rely on a human understanding of the underlying physical processes, our approach has the potential to lead to new insights into poorly understood many-body phenomena such as MBL or topological phases through an analysis of the feature extraction layer.
The tools for this are still in their infancy, but evolving at a fast pace.

In future work, we will apply the technique demonstrated here to larger system sizes, hope to gain new insights into whether there are additional thermodynamically stable phases near the MBL phase transition \cite{Kjaell2014,Goold2015}, and plan to fully explore the flexibility of the adversarial approach, that allows to suppress the dependence of the output on specific non-universal features.

\section*{Acknowledgments}
C.\ G.~acknowledges support by the European Union's Marie Sk\l{}odowska-Curie Individual Fellowships (IF-EF) programme under GA: 700140.  We also acknowledge financial support from the European Research Council
(CoG QITBOX and AdG OSYRIS), the Axa Chair in Quantum Information Science,
Spanish MINECO (FOQUS FIS2013-46768, QIBEQI FIS2016-80773-P and Severo Ochoa Grant No.~SEV-2015-0522, FisicaTeAMO FIS2016- 79508-P), EU STREP program EQuaM
(FP7/2007-2017, Grant No. 323714), EU grant QUIC (H2020- FETProAct-2014 641122), Fundaci\'{o} Privada Cellex, and Generalitat de Catalunya (Grant No.~SGR 874 and 875, and CERCA Programme). P.\ H. acknowledges that this project has received funding from the European Union’s Horizon 2020 research and innovation programme under the Marie Skłodowska-Curie grant agreement No 665884, as well as by the ‘Severo Ochoa 2016-2019' program at ICFO (SEV-2015-0522). A.\ D. is financed by a Cellex-ICFO-MPQ fellowship and by a Juan de la Cierva fellowship (IJCI-2017-33180).
We acknowledge a hardware donation by Nvidia Corporation.

\section*{Appendix}
\begin{figure*}[htp!]
    \centering
    \subfigure[]{	
        \includegraphics[width=0.3\linewidth]{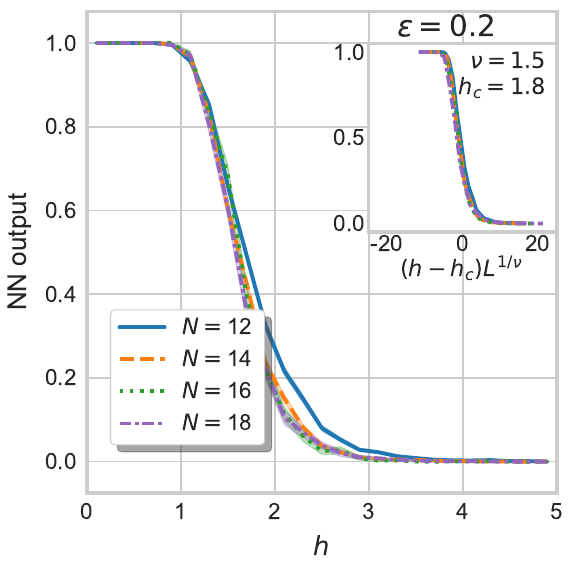}
        \label{fig:nn_output_eps02}
    }
    \subfigure[]{
        \includegraphics[width=0.3\linewidth]{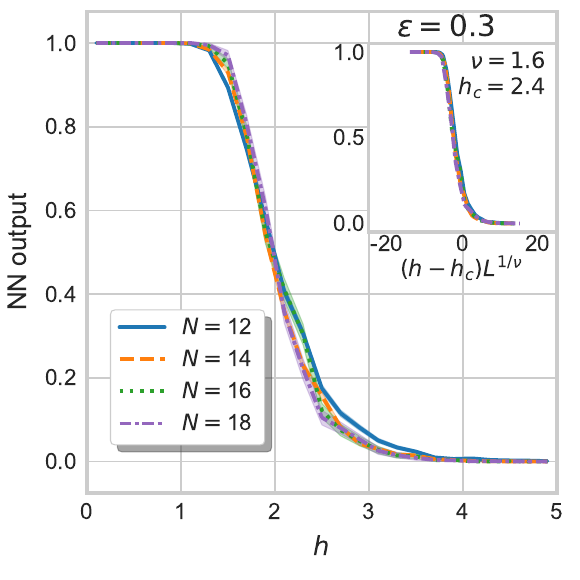}
        \label{fig:nn_out_eps03}
    }
    \subfigure[]{
        \includegraphics[width=0.3\linewidth]{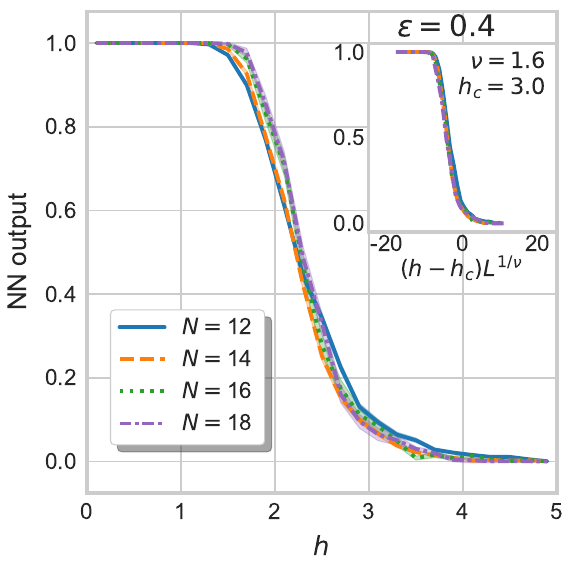}
        \label{fig:nn_out_eps04}
    }
    \subfigure[]{
        \includegraphics[width=0.3\linewidth]{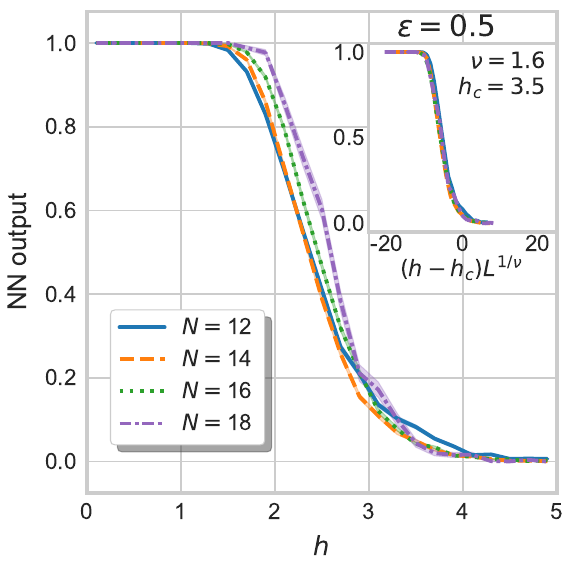}
        \label{fig:nn_out_eps05}
    }
    \subfigure[]{
        \includegraphics[width=0.3\linewidth]{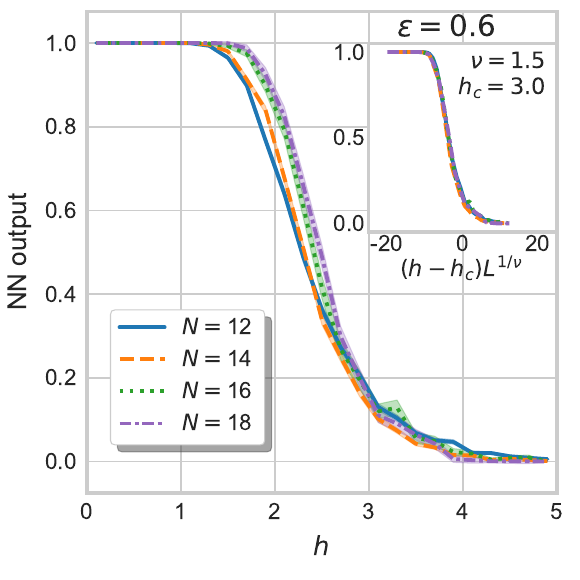}
        \label{fig:nn_out_eps06}
    }
    \subfigure[]{
        \includegraphics[width=0.3\linewidth]{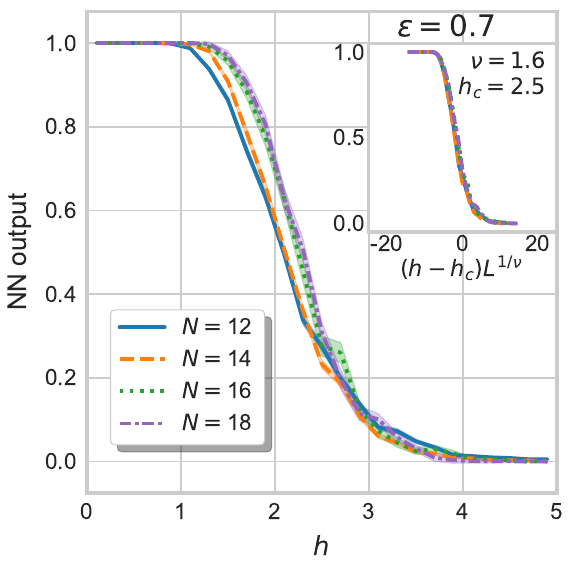}
        \label{fig:nn_out_eps07}
    }
    \subfigure[]{
        \includegraphics[width=0.3\linewidth]{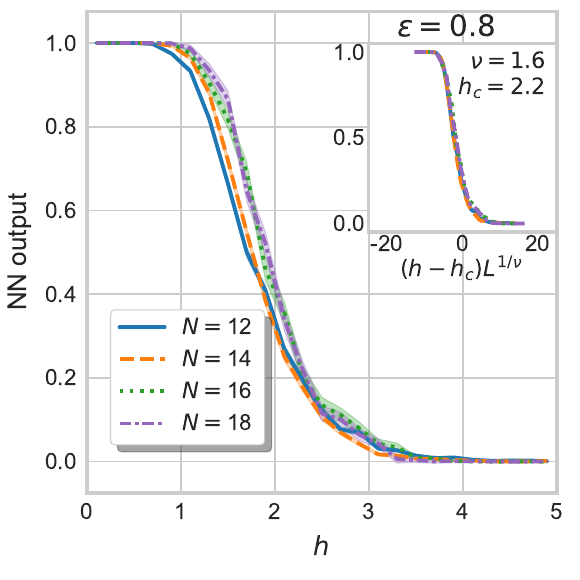}
        \label{fig:nn_out_eps08}
    }
    \caption{\label{fig:collapse_for_all_epsilon}
        Output of the neural network at energies $\epsilon = 0.2$ to $\epsilon = 0.8$. $N = 12$ and $14$ are averaged over $500$ disorder realizations and $N = 16$ and $18$ over $100$ realizations. The data collapse (inset) determines the position of the phase transition $h_c$ in the thermodynamic limit. The error bands show the ensemble standard deviation $s =  (\sum_i^N (x_i-\hat{x})^2 / (N-1))^{1/2}$ of the disorder average.}
\end{figure*}

\begin{table}[htb!]
    \centering
    \begin{tabular}{| c | c | c | c |}
        \hline
        $\epsilon$ & $\nu\pm\Delta \nu$ & $h_c\pm\Delta h_c$  \\ \hline
        $0.2$ & $1.5 \pm 0.2$ & $1.8 \pm 0.4$  \\ \hline
        $0.3$ & $1.6 \pm 0.2$ & $2.4 \pm 0.4$  \\ \hline
        $0.4$ & $1.6 \pm 0.2$ & $3.0 \pm 0.2$  \\ \hline
        $0.5$ & $1.6 \pm 0.1$ & $3.5 \pm 0.2$  \\ \hline
        $0.6$ & $1.5 \pm 0.2$ & $3.0 \pm 0.2$  \\ \hline
        $0.7$ & $1.6 \pm 0.2$ & $2.5 \pm 0.4$  \\ \hline
        $0.8$ & $1.6 \pm 0.2$ & $2.2 \pm 0.4$  \\ \hline
    \end{tabular}
    \caption{Estimates for $\nu$ and $h_c$ as well as their errors.
        The errors were conservatively estimated by fixing the best possible value for the respective other quantity and determining from plots such as those in Figure ~\ref{fig:collapse_for_all_epsilon} when the data collapse would diverge visibly.}
    \label{tab:phase-boundary-data}
\end{table}

\subsection{Details of the machine learning technique}

\subsubsection{Domain adversarial neural network}
\label{sec:domain_adversarial_neural_network}
The main idea behind a domain adversarial neural network \cite{ganin2016domain} is that we have two sets of input data.
The set of states deep inside the phase $\mathcal{D} = \{(x_i, y_i) \}$, which consists of pairs of data points $x_i$ and labels $y_i$ and the set of states $\mathcal{B} = \{ x_i \}$, which includes states close to the phase boundary and which is unlabeled.
The task of the neural network is to learn from the labeled instances and adapt this knowledge to the new unknown instances.
To achieve this, the domain adversarial neural network setup consists of three parts: the feature extractor, the phase discriminator, and the adversary.
The feature extractor filters the information from the input data, the phase discriminator classifies the state into the correct phase and the adversarial tries to distinguish labeled from unlabeled data instances, i.e., it tries to distinguish between the two sets $\mathcal{D}$ and $\mathcal{B}$.

The first part of the DANN, the feature extractor, consists of convolutional neural networks which map the input data to a high dimensional, abstract feature vector $\bm{f} = G_f( \bm{x} , \theta_f)$.
This latent representation of the state vectors is forwarded to the phase discriminator $\bm{d} = G_d( \bm{f} , \theta_d)$ and the adversarial $\bm{a} = G_a( \bm{f} , \theta_a)$.
The $\theta_i$ represent the parameters that have to be learned through the training.

Since phase labels are only given for the input data coming from $\mathcal{D}$, the phase discriminator loss $L_d$ is calculated on this set alone. 
The loss measures the binary cross-entropy between the network's outputs and the actual labels.
The adversarial loss $L_a$ can be calculated on all states $x \in \mathcal{D} \cup \mathcal{B}$.
The loss function is again the binary cross-entropy, but calculated on which set the instance comes from.
The crucial point about a DANN is that the feature representation $\bm{f}$ has to be invariant for both sets, that is, by looking at the last layer of the feature extractor, one cannot tell which set an instance comes from.

This means that the feature extractor should only extract features that are decisive to predict the correct phase label, but not the correct adversarial label.
This can be achieved by optimizing $E(\theta_f, \theta_d, \theta_a) = L_d(\theta_f, \theta_d)- L_a(\theta_f, \theta_a)$ and finding the saddle point $(\theta_f, \theta_d) = \underset{\theta_f, \theta_d}{\mathrm{argmin}}~ E( \theta_f, \theta_d, \theta_a)$ and $(\theta_a) = \underset{\theta_a}{\mathrm{argmax}} ~ E( \theta_f, \theta_d, \theta_a )$.
According to this optimization problem, the update rule for the feature extractor has the form $\Delta\theta_f = \mu \left(\frac{\partial L_d}{\partial \theta_f} - \frac{\partial L_a}{\partial \theta_f} \right)$, which can be implemented via standard stochastic gradient descent and a gradient reversal layer between the feature extractor and the domain classifier.
This kind of training leads to a feature representation based on which the adversary cannot classify correctly because the neural network is unable to tell anymore which of the two sets $\mathcal{D}$ and $\mathcal{B}$ state comes from.
The phase discriminator at the same time, has learned invariant features of the input states to tell which phase it comes from.

\subsubsection{Details of the neural network architecture}
\label{sec:neural_network_computational_details}
The feature extractor of our DANN consists of four one-dimensional convolutional layers with four filters each and a filter length of three.
This means that a single filter in a single layer has a sliding window of length three that is convoluted with the input to the layer.
The input to the first layer is the ground state.
Each layer has four of these filters, extracting different features of the input to the layer.
The activation function of these layers are rectified linear units (ReLUs).
This is a piecewise linear function that outputs zero for negative values and a linear response for positive values.
While neural networks traditionally used nonlinear activation functions, the ReLUs have better numerical properties when training the network with many layers.
Each of those layers is followed by a max-pooling layer that pools from three neighboring neurons.
This is a critical step for coarse-graining the representation: we pick the maximum of the value of the activation over three neighboring points and discard the other two.
In effect, we reduce the dimension of the vector by two-thirds in each of these pooling layers.
Pooling does not only ensure a lower-dimensional representation, but it also enables that the subsequent convolutional layer identifies longer range correlations in the original data.

Both the phase discriminator and adversarial networks have a single hidden layer with 128 neurons that are fully connected and activated by a ReLU function.
The final output of both fully connected networks consists of two neurons with a softmax activation function.
The softmax activation exponentiates each output and normalizes it with the partition function over the output layer.
In other words, we get a normalized probability distribution as the final outcome, which has important physical meaning in our work.
We apply batch normalization after every layer, which introduces a slight stochastic variation in the scale of the characteristics of the input states, and thus reduces the chance of overfitting.
Furthermore, we use dropout \cite{srivastava2014dropout} for the fully connected layers in the phase discriminator and adversarial networks, which is standard practice in achieving better performance.

\subsection{Hardware}
The numerical experiments were run on an Intel Xeon E5-1650 v4 with six physical cores clocked at 3.60~GHz base frequency and with 128~GByte of RAM.
The CPU was complemented by a Tesla K40 graphics processing units for training the neural networks, with 2880 physical cores clocked at 745~MHz base frequency and 12~GByte of GRAM.

\subsection{Input data}
\label{sec:input_data}
The set of states from deep inside the phases $\mathcal{D}$ was drawn from $h \in [0.1, 0.5]$  for the delocalized phase and from $h \in [7.0, 8.0]$ with a step size $0.1$ for the MBL phase for energy densities in the range $\epsilon \in [0.05, 0.95]$.
To have equally big sets in the delocalized and the MBL phase the values of $h$ are separated by steps of $0.05$ in the delocalized phase, $0.1$ respectively in the MBL phase. The epsilon values are separated by steps of $0.05$.
For the set close to the phase boundary we choose states with disorder strength $h \in [0.5, 7.0]$ separated by steps of $0.2$ and normalized energy $\epsilon \in [0.05, 0.95]$ in steps of $0.05$. 
For each set of parameters and disorder realization we find the 50 states closest to the chosen energy density $\epsilon$.
We take several realizations for each point in the parameter space which is chosen such that both sets are of the same dimension, namely 50k.
We have checked that the results do not depend on the details of how these sets are chosen.
The states were produced with the open-source software from~\cite{Pietracaprina2018}.

\subsection{Data analysis}
We analyze the data generated by the neural network in the way described in the main text.
In particular, to obtain an estimate of the energy resolved phase boundary in the thermodynamic limit, we compute the output of the phase classifier as a function of $h$ for various values of $\epsilon$ and then perform a data collapse as described in the main text.
The raw and collapsed data for different values of $\epsilon$ is shown in Figure~\ref{fig:collapse_for_all_epsilon}.
The resulting estimates for $h$, $\nu$, and their errors are shown in Table~\ref{tab:phase-boundary-data}.

\end{document}